\documentclass[conference]{IEEEtran}
\IEEEoverridecommandlockouts
\usepackage{amsmath,amssymb,amsfonts}
\usepackage{algorithmic}
\usepackage{graphicx}
\usepackage{subcaption}
\usepackage[dvipsnames]{xcolor}
\usepackage{textcomp}
\usepackage{verbatim}
\usepackage{xspace}
\usepackage{fancyvrb}
\usepackage{fvextra}
\usepackage{inconsolata}
\usepackage{colortbl}
\usepackage{url}
\usepackage{mathptmx}
\usepackage{wrapfig}
\usepackage{chato-notes}
\usepackage{hyperref}
\usepackage[capitalize]{cleveref}
\usepackage{placeins}
\usepackage{multirow}
\usepackage{booktabs}
\usepackage{float}
\usepackage{breqn}
\usepackage{csquotes}
\usepackage[inline]{enumitem}

\usepackage[backend=biber, minbibnames=1, maxbibnames=20, bibstyle=ieee, citestyle=numeric-comp, mincitenames=1, maxcitenames=2]{biblatex}
\usepackage[english]{babel}
\AtEveryBibitem{\clearfield{pages}}
\everymath{\displaystyle}

\crefname{section}{Sect.}{Sect.}
\Crefname{section}{Sect.}{Sect.}
\crefname{figure}{Fig.}{Fig.}
\Crefname{figure}{Fig.}{Fig.}
\crefname{table}{Table}{Table}
\Crefname{table}{Table}{Table}

\newcommand{\pptext}[1]{\smallskip\noindent \textit{#1.}\xspace}
\newcommand{\example}{\pptext{Example}} 
\newcommand{\cd}[1]{\texttt{\small #1}\xspace}
\newcommand{\fncd}[1]{\texttt{\footnotesize #1}\xspace} 

\newcommand{\edgewise}{\xspace{\small\sc EdgeWise}\xspace}
\newcommand{\edgewisecr}{\xspace{\small\sc EdgeWiseCR}\xspace}
\newcommand{\plnot}{{\scriptsize\textbackslash}+}

\DefineVerbatimEnvironment{code}{Verbatim}
{gobble=4, fontfamily=zi4, fontsize=\footnotesize, frame=single, framesep=1mm, framerule=0.1pt, rulecolor=\color{gray}}

\DefineVerbatimEnvironment{codeNum}{Verbatim}
{gobble=4, fontfamily=zi4, numbers=left, numbersep=5pt, numberblanklines=false, firstnumber=last, tabsize=2, fontsize=\footnotesize, frame=single, framesep=1mm, framerule=0.1pt, rulecolor=\color{gray}, commandchars=\\\{\}}

\def\BibTeX{{\rm B\kern-.05em{\sc i\kern-.025em b}\kern-.08em
    T\kern-.1667em\lower.7ex\hbox{E}\kern-.125emX}}

\begin{document}

\title{Combining Declarative and Linear Programming for Application Management in the Cloud-Edge Continuum\thanks{Work partly funded by projects: \textit{Energy-aware management of software applications in Cloud-IoT ecosystems} (RIC2021PON\_A18), funded with ESF REACT-EU resources by the \textit{Italian Ministry of University and Research} through the \textit{PON Ricerca e Innovazione 2014--20} and \textit{hOlistic Sustainable Management of distributed softWARE systems (OSMWARE)}, UNIPI PRA\_2022\_64, funded by the University of Pisa, Italy.}
}

\author{
 \IEEEauthorblockN{Jacopo Massa, Stefano Forti, Patrizio Dazzi, Antonio Brogi}
 \IEEEauthorblockA{\textit{Department of Computer Science} \\
 \textit{University of Pisa}\\
 Pisa, Italy }
}

\maketitle

\begin{abstract}
This work investigates the data-aware multi-service application placement problem in Cloud-Edge settings. We previously introduced \edgewise, a hybrid approach that combines declarative programming with Mixed-Integer Linear Programming (MILP) to determine optimal placements that minimise operational costs and unnecessary data transfers. The declarative stage pre-processes infrastructure constraints to improve the efficiency of the MILP solver, achieving optimal placements in terms of operational costs, with significantly reduced execution times.
In this extended version, we improve the declarative stage with continuous reasoning, presenting \edgewisecr, which enables the system to reuse existing placements and reduce unnecessary recomputation and service migrations. In addition, we conducted an expanded experimental evaluation considering multiple applications, diverse network topologies, and large-scale infrastructures with dynamic failures. The results show that \edgewisecr achieves up to 65\% faster execution compared to \edgewise, while preserving placement stability under dynamic conditions.

\end{abstract}

\begin{IEEEkeywords}
application placement, data-awareness, mathematical optimisation, logic programming, continuous reasoning
\end{IEEEkeywords}

\section{Introduction}
\label{sec:introduction}

Cloud-Edge computing has emerged as a crucial paradigm to accommodate the ever-growing need for computational power and data storage. This approach addresses latency issues and boosts resource efficiency by distributing workloads across a spectrum of resources, ranging from centralised Cloud data centres to peripheral Edge nodes~\cite{duan2020convergence}. One of its primary benefits is mitigating the \textit{data deluge}~\cite{tortonesi2019taming} caused by the rapid expansion of the Internet of Things (IoT), ensuring that computational, networking and storage services are brought closer to both data sources and consumers. However, efficiently mapping multi-service applications to this distributed infrastructure while satisfying both functional and non-functional constraints remains a significant challenge~\cite{salaht2020overview}.

Traditional methods for service placement can be broadly classified into heuristic-based approaches and mathematical optimisation models. Heuristic methods provide quick approximations, but often fsail to identify optimal placements, especially in complex and dynamic environments. Conversely, Mixed-Integer Linear Programming (MILP) techniques offer mathematically optimal solutions but at the expense of substantial computational overhead and extended execution times~\cite{brogi2019place}. More recently, declarative programming approaches~\cite{forti2022declarative,massa2022daplacer} have been explored for their ability to handle qualitative constraints, such as security requirements and data locality policies, in a more flexible and extensible way. However, declarative methods alone may struggle to balance computational efficiency with placement quality.

Our prior work introduced a hybrid methodology that integrates declarative programming with MILP-based optimisation to facilitate optimal placement of multi-component applications, made of both services and functions, while reducing execution overhead~\cite{edgewise}. In that approach, the declarative stage is a preliminary filtering mechanism, eliminating infeasible placement candidates based on infrastructure constraints and service dependencies. We adopted the perspective of an infrastructure provider managing the Cloud-Edge continuum. Therefore, the cost associated with a placement is interpreted as the provisioning cost for deploying application components across the available infrastructure. Our method significantly improves computational efficiency by reducing the complexity of the problem before the MILP solver is engaged. Additionally, the model incorporates data-awareness by evaluating communication costs and inter-service dependencies, ensuring that placements are optimised for computational feasibility and overall system performance. Through this strategy, our hybrid approach achieved up to $10\times$ reduction in execution time compared to standalone MILP solutions, all while maintaining placement optimality.

Compared to our previous work~\cite{edgewise}, this extended version introduces the following key contributions:
\begin{itemize}
    \item we enhanced declarative preprocessing with a continuous reasoning~\cite{forti2022declarative,massa2022daplacer} mechanism that tries to replace only part of the application components, reducing redundant computations and unnecessary service migrations.
    
    \item we released a new and extended open-source implementation of the enriched hybrid pipeline, \edgewisecr, including the declarative and MILP stages, now enriched with continuous reasoning support.
    
    \item we conducted a more comprehensive experimental study involving three realistic application scenarios, larger infrastructures (up to 2048 nodes), and a simulation environment with dynamic failures and placement updates.
\end{itemize}

The findings indicate that \edgewisecr executes up to 65\% quicker compared to our former tool \edgewise~\cite{edgewise}, with a cost increase of up to 33\%, all while maintaining stable placement in dynamic environments.

The rest of this article is organised as follows. \Cref{sec:problem} introduces the motivating use case and the formalisation of the considered problem. \Cref{sec:formulation} recalls the original declarative and MILP formulations. \Cref{sec:prolog_preprocessing} presents our enhanced pipeline, detailing the continuous reasoning mechanism. \Cref{sec:experiments} reports on the experimental setup and the evaluation of our extended approach. Finally, \cref{sec:related_work} discusses related work and \cref{sec:conclusions} concludes by pointing out some directions for future work.
\section{Use Case and Considered Problem}
\label{sec:problem}

\begin{figure}[t]
    \centering
    \includegraphics[width=.48\textwidth, trim={0cm 0cm 0cm 1.5cm}, clip]{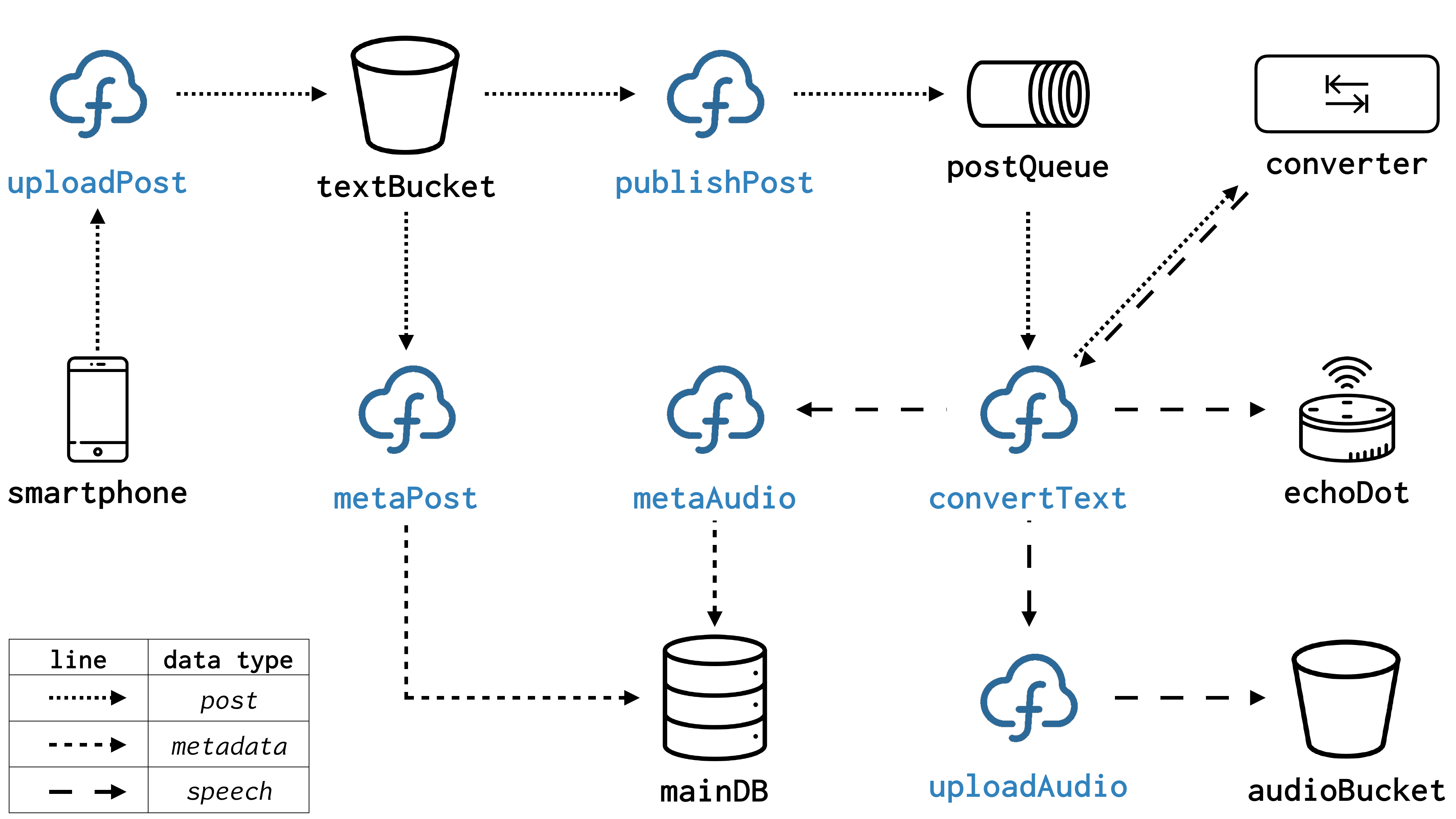}
    \caption{Example application.}
    \label{fig:app_example}
\end{figure}

The \cd{speakToMe} application, depicted in \cref{fig:app_example}, offers a modular text-to-speech service, allowing users to submit short text posts that are translated into audio messages and made available for playback in multiple languages across different devices. The application is built as a Cloud-Edge pipeline, comprising six functions and five services, each tailored to specific computational and operational requirements, and responsible for a distinct processing phase.

The workflow begins when a user submits a text post through its device, which is processed by the \cd{uploadPost} function. The post content is saved in the \cd{textBucket} storage service, while the \cd{metaPost} function extracts and stores its metadata in the central database \cd{mainDB}.
Once stored, the post enters a processing queue managed by the \cd{postQueue} service. The content is then retrieved and passed to the \cd{convertText} function, which performs translation into the target language. The translated text is forwarded to the \cd{converter} service, which prepares it for audio synthesis. The resulting audio is stored in the \cd{audioBucket} service through the \cd{uploadAudio} function. Finally, the \cd{metaAudio} function extracts relevant metadata from the audio and stores it again in the \cd{mainDB}.
This pipeline ensures that users receive high-quality, language-specific audio representations of their input posts, while enabling efficient metadata management, scalable storage, and optimised processing across Cloud and Edge nodes. 

Services handle long-running, resource-intensive tasks, such as translation processing and speech synthesis. At the same time, functions are executed on demand in response to specific events, such as user input submission or query retrieval. For example, the \cd{converter} service requires significant computational power due to real-time audio synthesis. In contrast, the \cd{metaPost} function primarily manages user content metadata with minimal processing overhead.
A key distinction between services and functions is their scope and execution model. Services typically operate continuously in the background, managing multiple interdependent tasks, while functions are stateless and triggered only when needed, executing briefly before terminating. This differentiation is crucial in Cloud-Edge computing environments, where balancing persistent services with lightweight function executions enhances scalability and resource allocation.
In addition, each function instance maintains metadata regarding request frequency, processing time, and computational resource usage.
\Cref{tab:requirements} lists the above information for our use case. For example, \cd{uploadFun} requires a \cd{python} platform executed on an \cd{x86} architecture with at least \cd{20} units of available hardware. 
The \cd{uploadPost} instance of this type of function is predicted to receive \cd{1000} requests per month, each lasting \cd{30ms}.

Services and functions continuously exchange data with each other and with IoT devices. As shown in \cref{fig:app_example}, these flows vary in data size, security requirements, latency constraints, and data rates, depending on the type and direction of interaction.
For instance, the \cd{uploadPost} function and the \cd{textBucket} service exchange the \cd{0.4MB}-unit \cd{post} data at a rate of \cd{5Hz}, with a maximum data transfer time of \cd{60ms}, which requires an authentication service and encrypted storage.

Furthermore, consider a Cloud-IoT infrastructure made of heterogeneous computing \textit{nodes}, interconnected via wireless and wired \textit{links}. The nodes are characterised by \textit{software}, \textit{hardware}, \textit{security} and additionally \textit{IoT} capabilities. Featured \textit{latency} and \textit{bandwidth} describe physical links between the nodes.

\begin{table*}[ht]
\centering
\resizebox{0.65\textwidth}{!}{%
\begin{tabular}{|c|c|c|c|cl|}
\hline
\textbf{Component} & \textbf{SWReqs} & \textbf{Arch} & \textbf{HWReqs} & \multicolumn{2}{c|}{\textbf{\begin{tabular}[c]{@{}c@{}}Instances\\ (MonthlyRequests, TimeDuration)\end{tabular}}} \\ \hline
storage & ubuntu & x86 & 100 & \multicolumn{2}{c|}{} \\ \hline
database & mySQL, ubuntu & x86 & 50 & \multicolumn{2}{c|}{} \\ \hline
queue & python & x86 & 2 & \multicolumn{2}{c|}{} \\ \hline
textToSpeech & gcc, python & arm64 & 4 & \multicolumn{2}{c|}{} \\ \hline
\multirow{2}{*}{uploadFun} & \multirow{2}{*}{python} & \multirow{2}{*}{x86} & \multirow{2}{*}{20} & \multicolumn{2}{c|}{uploadPost (1000, 30)} \\ \cline{5-6} 
 &  &  &  & \multicolumn{2}{c|}{uploadAudio (1000, 20)} \\ \hline
\multirow{2}{*}{metadataFun} & \multirow{2}{*}{python} & \multirow{2}{*}{arm64} & \multirow{2}{*}{2} & \multicolumn{2}{c|}{metaPost (1500, 8)} \\ \cline{5-6} 
 &  &  &  & \multicolumn{2}{c|}{metaAudio (2500, 130)} \\ \hline
publishFun & js & x86 & 4 & \multicolumn{2}{c|}{publishPost (2000, 8)} \\ \hline
ttsFun & python & arm64 & 30 & \multicolumn{2}{c|}{convertTxt (2500, 30)} \\ \hline
\end{tabular}%
}
\smallskip
\caption{Application components' requirements.}
\label{tab:requirements}
\end{table*}

\subsection{Considered problem}
\noindent Application management needs to decide on the optimal node(s) for deploying application components (services and functions) while minimising the application part involved in runtime reallocation. Subsequently, we will explore two methods for modelling and addressing this issue:
\smallskip

\fbox{
\centering
\begin{minipage}{0.42\textwidth}
Find a placement of a multi-service application onto a Cloud-IoT infrastructure, viz., a complete mapping $P$ from application components to Cloud-IoT nodes, that 
    \begin{enumerate*}[label=\textit{(\roman{*})}]
        \item meets all application hardware, software, IoT and data security, latency and bandwidth requirements, 
        \item minimises provisioning costs and 
        \item reduces the number of service migrations w.r.t. a previously enforced placement.
    \end{enumerate*}
\end{minipage}
}


\section{Declarative \& MILP formulation}
\label{sec:formulation}

This section begins with an overview of the Prolog programming language (\cref{sec:prolog_background}). We then offer a full declarative formulation of the considered problem with Prolog in \cref{sec:prolog_formulation}. Subsequently, the MILP formulation is discussed in \cref{sec:milp_formulation}. \cref{sec:prolog_preprocessing} and \cref{sec:pipeline-overview} report on the pre-processing steps and the overall structure of the pipeline, respectively.

\subsection{Background: Prolog}
\label{sec:prolog_background}
Prolog programs consist of \textit{clauses} structured as ``\cd{a :- b1, ..., bn.}'' which means: \cd{a} holds if \cd{b1} $\wedge \ldots \wedge$ \cd{bn} holds. Clauses with empty premises (\cd{n}$=0$) are \textit{facts}. Predicate definitions can also include \textit{disjunctions} (denoted by ``\cd{;}'') and negations (denoted by ``\cd{\textbackslash+}''). Variables begin with uppercase letters and lists are represented using square brackets (e.g. \cd{[L|Ls]} where \cd{L} is the head and \cd{Ls} the tail). A \textit{predicate} in Prolog is denoted as \cd{predicateName/arity}, where \cd{predicateName} is the name and \cd{arity} is the number of arguments it accepts.
For example, \cd{city/2} indicates a predicate \cd{city} with two arguments.
Prolog programs can be queried, and the Prolog interpreter tries to answer each query by applying \textit{Selective Linear Definite} (SLD) resolution~\cite{lloyd} and returning a computer answer substitution that instantiates the variables in the query. In this article, we rely on the SWI-Prolog implementation~\cite{swipl}. For example, the query \cd{?- romantic(City).} on the program

\begin{code}
    city(paris). 
    city(pisa). 
    city(london).
    
    rainy(pisa). closeToTheSea(pisa).
    
    romantic(City):- 
        city(City), 
        (rainy(City), closeToTheSea(City)) ; (City = paris).
\end{code}

\noindent
returns the computed answer substitutions

\begin{code}
    ?- romantic(City).
    City = pisa ; City = paris.
\end{code}

\noindent Rewriting the query by applying the first clause to retrieve a city non-deterministically, the second and third to check whether the retrieved city is rainy and close to the sea or it is Paris. This also shows the \textit{closed-world} assumption of Prolog. In fact, we can prove that London is not romantic according to the knowledge base provided and the definition of \cd{romantic/1}.

\subsection{Declarative model formulation}
\label{sec:prolog_formulation}

\pptext{Applications} Next-gen applications \cd{App} are described in terms of their set of \cd{Services} and \cd{Functions} as in 
\smallskip
\begin{code}
    application(App, Functions, Services).
\end{code}

Each service \cd{S} is defined with its required software stack \cd{SWReqs} and execution constraints, which include hardware \cd{Architecture} (e.g. \cd{arm64}, \cd{x86}) and computational resources \cd{HWReqs}\footnote{For the sake of simplicity, we represent generic hardware units as positive numbers as most of the approaches surveyed in \cite{brogi2019place}.} necessary to run \cd{S}:

\begin{code}
    service(S, SWReqs, (Architecture, HWReqs)).
\end{code}

Similarly, functions \cd{F} are declared with the software environment \cd{SWPlat} they rely on, and a pair containing the \cd{Architecture} and the hardware requirements \cd{HWReqs}, as in

\begin{code}
    function(F, SWPlat, (Architecture, HWReqs)).
\end{code}

IoT devices \cd{T}, which provide data and interact with application components, are described by their \cd{Type} (e.g., \cd{thermostat}, \cd{firealarm}) as in

\begin{code}
    thing(T, Type).
\end{code}

Instances of these components (i.e., \cd{services}, \cd{functions} and \cd{things}) represent concrete deployments, establishing links between abstract service and function definitions and their real-world instantiations:
\smallskip
\begin{code}
    serviceInstance(SIId, S).
    functionInstance(FIId, F, (MonthlyRequests, TimeDuration)).
    thingInstance(TIId, T).
\end{code}

\noindent Note that function instances also specify the expected number of \cd{MonthlyRequests} and the average \cd{TimeDuration} of such requests expressed in milliseconds.

The above instances are composed into applications by means of data flows like
\smallskip
\begin{code}
    dataFlow(Src, Dst, DataType, SecReqs, Size, Rate, MaxLatency).
\end{code}

\noindent 
where \cd{Src} and \cd{Dst} are the instance identifiers of communicating application components that exchange information of a certain \cd{DataType}, requiring a list of security requirements \cd{SecReqs} at the \cd{Src} and \cd{Dst} deployment nodes, featuring a certain data \cd{Size} (in MB) and transmission \cd{Rate} (in Hz), and tolerating a data transfer time of at most \cd{MaxLatency}.

\example For instance, a portion of the application described in \cref{sec:problem} and sketched in \cref{fig:app_example} can be declared as:
\smallskip
\begin{code}[samepage=true]
    service(storage, [ubuntu], (x86, 100)).
    function(uploadFun, python, (x86, 20)).
    thing(smph, smartphone).
    
    application(speakToMe, [uploadPost, ...], [textBucket, ...]).
    serviceInstance(textBucket, storage).
    functionInstance(uploadPost, uploadFun, (1000, 30)).
    thingInstance(iphoneXS, smph).
    
    dataFlow(uploadPost, textBucket, post, 
             [authentication, enc_storage], 0.4, 5, 80).
\end{code}

\pptext{Infrastructure} The infrastructure nodes are represented by a set of facts such as
\smallskip
\begin{code}
    node(N, SWCaps, (Architecture, HWCaps), SecCaps, IoTDevices).
    nodeType(N, Type).
    location(N, Location).
    provider(N, Prodiver).
\end{code}

\noindent
with the node \cd{Type} (e.g. \cd{cloud}, \cd{edge} or \cd{thing}), the node \cd{Location} (e.g., \cd{it}, \cd{es}, \cd{fr}), the node \cd{Provider} (e.g. \cd{aws}, \cd{azure}), the list of available software \cd{SWCaps}, a pair containing the node \cd{Architecture} and free hardware \cd{HWCaps}, the featured security capabilities \cd{SecCaps} and the hosted \cd{IoTDevices}. 

\noindent Finally, end-to-end links between nodes \cd{N1} and \cd{N2} are described by their average \cd{Latency} and \cd{Bandwidth}, as facts like

\begin{code}
    link(N1, N2, Latency, Bandwidth).
\end{code}

\noindent
Note that this infrastructure representation allows expressing further requirements (e.g., geo-location, custom security policies, providers' affinity) for a service or function \cd{C} onto a certain \cd{Node} via a predicate like

\begin{code}
    requirements(C, Node) :- ...
\end{code}

\pptext{Cost model}
The \cd{Cost} of using a certain \cd{NodeType} for deploying a component \cd{C}, interpreted as its provisioning cost, slightly varies depending on whether \cd{C} is a service or a function, but is always computed via a predicate like

\begin{code}
    cost(NodeType, C, Cost) :- ...
\end{code}


For \textit{services}, the cost computation aggregates the required software price and the assigned hardware's computational expense. The \cd{unitCost/3} predicate defines unit prices for different architectures (e.g., \cd{arm64}, \cd{x86}) and software dependencies (e.g., \cd{ubuntu}, \cd{python})\footnote{Unit costs are inspired by the AWS EC2 pricing: \url{https://aws.amazon.com/ec2/pricing/}}. The overall cost is computed by multiplying the hardware requirements by the corresponding unit price and summing up software-related costs:

\begin{code}[fontsize=\scriptsize]
    cost(NType, SIId, C) :-
        serviceInstance(SIId, SId),
        service(SId, SWReqs, (Arch, HW)),
        findall(C1, (unitCost(S, NType, C1), member(S, SWReqs)), SWCosts),
        sum_list(SWCosts, SWCost),
        unitCost(Arch, NType, HWC),
        HWCost is HW * HWC,
        C is HWCost + SWCost.
\end{code}

For \textit{functions}, the cost model accounts for both computational and request-processing costs\footnote{Computational costs and request-processing costs are inspired by the AWS Lambda pricing: \url{https://aws.amazon.com/lambda/pricing/}}, varying by deployment on different \cd{NodeType}s. Computational costs depend on the function's processing power, number of requests per month, and request duration. The pricing structure differs across infrastructures, reflecting variations in energy efficiency and operational costs:

\begin{code}[fontsize=\scriptsize]
    cost(cloud, FIId, C) :-
        functionInstance(FIId, FId, (ReqXMonth, ReqDuration)),
        function(FId, _, (_, HWReqs)),
        Mbps is HWReqs * ReqXMonth * ReqDuration / 1000, 
        CompCost is Mbps * <UnitCompCost>,
        ReqCost is ReqXMonth * <UnitReqCost>,
        C is CompCost + ReqCost.
\end{code}

Similarly, different pricing applies to \cd{edge} and \cd{thing} deployments, changing \cd{UnitCompCost} and \cd{unitReqCost} accordingly.

\example For instance, consider a simple infrastructure consisting of two nodes:
\begin{itemize}
    \item a 200-unit hardware x86 architecture Cloud node hosted in Italy by Microsoft Azure, providing firewall and access logs services, supporting python and javascript, and
    \item a 50-unit hardware arm64 architecture Edge node located in Spain by Amazon Web Services and linked to a smartphone, providing wireless security and anti-tampering services, and support to the C language and MySQL.
\end{itemize}
    
These two nodes communicate through a bidirectional link that features \cd{10ms} latency and \cd{150Mbps} bandwidth. This infrastructure can be declared as:
\smallskip

\begin{code}[samepage=true]
    node(n1, [python, js], (x86, 200), 
         [access_logs, firewall] []).
    node(n2, [gcc, mySQL], (arm64, 50), 
         [wireless_security, anti_tampering], [iphoneXS]).
    
    link(n1, n2, 10, 150).
    link(n2, n1, 10, 150).
    
    nodeType(n1, cloud).
    nodeType(n2, edge).
    
    location(n1, it).
    location(n2, es).
    
    provider(n1, azure).
    provider(n2, aws). 
\end{code}

\noindent Retaking the \cd{database} component of the application described in \cref{sec:problem}, the application provider can narrow the selection of eligible nodes, requiring that the node provider must be Azure or AWS, and it must provide secure storage enhanced by access logs, authentication services, and a total ingress bandwidth of at least \cd{200Mbps}. Such policies and requirements can be declared as:
\smallskip

\begin{code}[samepage=true]
    requirements(database, N) :-
      node(N, _, _, SecCaps, _),
      (provider(N, 'aws'); provider(N, 'azure')),
      secure_storage(SecCaps),
      member(access_logs, SecCaps), 
      member(authentication, SecCaps),
      avgInBW(N, 200).
    
    secure_storage(SC) :- 
      member(backup, SC);
      (member(enc_storage, SC), member(obfuscated_storage, SC)).
\end{code}

\subsection{MILP prototype}
\label{sec:milp_formulation}
We now illustrate a MILP solution to a sub-problem of the one described in \cref{sec:problem}, which only takes into account numerical constraints, in particular the cumulative requirements and capabilities of \textit{hardware} and \textit{bandwidth}, but also the requested and featured \textit{latency}. Our notation is summarised in \cref{tab:notation}.

\begin{table}[ht]
\centering
{\resizebox{0.95\columnwidth}{!}{%
\begin{tabular}{|c|c|}
\hline
\textbf{Notation} & \textbf{Description}                                             \\ \hline
$S$                 & list of the application components (i.e. functions and services) \\ \hline
$N$                 & list of infrastructure nodes                                     \\ \hline
$MAX\_BIN$         & maximum number of nodes that can be used                         \\ \hline
$x_{ij}$          & component \textit{i} is placed on node \textit{j}         \\ \hline
$c_{ij}$          & cost of placing component \textit{i} on node \textit{j}          \\ \hline
$b_{j}$           & node \textit{j} is used in the placement          \\ \hline
$hwTh$, $bwTh$        & hardware and bandwidth thresholds                                \\ \hline
$rhw_{i}$         & hardware requirements of component \textit{i}                    \\ \hline
$fhw_{j}$         & hardware capabilities of node \textit{j}                         \\ \hline
$rbw_{ih}$        & bandwidth requirements of data flow between components \textit{i}, \textit{h}\\ \hline
$fbw_{jk}$        & bandwidth capabilities of link between nodes \textit{j}, \textit{k}  \\ \hline
$rlat_{ih}$        & latency requirements of data flow between components \textit{i}, \textit{h}\\ \hline
$flat_{jk}$        & latency capabilities of link between nodes \textit{j}, \textit{k}  \\ \hline
\end{tabular}
}\smallskip
\caption{MILP model notation dictionary.}
\label{tab:notation}
}
\end{table}

\pptext{Problem formulation} Given an application made of a set of components \textit{S} and an infrastructure made of a set of nodes \textit{N}, we want to assign every component in \textit{S} to a node in \textit{N}. 

Let \textit{X} be the matrix that assigns each component \textit{s} to the node \textit{n} that will host it, the binary variable $ x_{ij} = 1$ indicates that the component $s_i$ is hosted on node $n_j$, otherwise $x_{ij} = 0$. The matrix \textit{C} contains the coefficients $c_{ij}$ that quantify the cost of placing $s_i$ on $n_j$. This formulation leads to the following optimisation problem:

\begin{equation}
    \label{eq:objective_function}
    \min \sum_{i,j} c_{ij} \cdot x_{ij}
\end{equation}

\pptext{Constraints} To be aligned with Prolog modelling (\cref{sec:prolog_formulation}) we must consider the following constraints:
\begin{itemize}
    \item Each $s_i$ must be assigned to only one $n_j$. Thus, the sum of values of each row in \textit{X} must be equal to 1:
        \begin{equation}
        \begin{array}{rcr}
            \sum_{j \in N} x_{ij} = 1 & & \forall i \in S
        \end{array}
        \end{equation}
    \item We impose a maximum number of nodes (i.e. \cd{MAX\_BIN}) that can be used for the placement, tracking the used nodes with a set of binary variables \textit{B}:
        \begin{equation}
        \begin{array}{rrcl}
            \forall i \in S, j \in N: & x_{ij} & \leqslant & b_j \\
            \forall j \in N: & \sum_{j}b_{j} & \leqslant & MAX\_BIN
        \end{array}
        \end{equation}
        
    \item The total amount of \textit{hardware} required by each component $s_i$ (i.e. $rhw_i$) on a node $n_j$ must be lower than its free hardware capability $fhw_j$, also considering the hardware infrastructure threshold \cd{hwTh}, which accounts for a reserved amount of hardware to avoid node overloading:
        \begin{equation}
        \begin{array}{cc}
            \forall j \in N: &\sum_{i}(x_{ij} \cdot rhw_i) \leqslant (fhw_j - hwTh) \\
        \end{array}
        \end{equation}

    \item The \textit{latency} featured by each link between nodes $n_j$ and $n_k$ (i.e $flat_{jk}$) must be lower than the latency requested by each data flow between components $s_i$ and $s_h$ (i.e. $rlat_{ih}$):
        \begin{equation}
        \label{eq:lat}
        \begin{array}{cc}
            \forall i,h \in S, \forall j,k \in N & x_{ij} \cdot x_{hk} \cdot flat_{jk} \leqslant rlat_{ih} \\
        \end{array}
        \end{equation}
    
    \item The total amount of \textit{bandwidth} required by data exchanged between pair of components $s_i$ and $s_h$ (i.e. $rbw_{ih}$), flowing through the link between nodes $n_j$ and $n_k$ must be lower than the bandwidth featured by the link $fbw_{jk}$, considering also the bandwidth infrastructure threshold \cd{bwTh}, which accounts for a reserved amount of bandwidth to avoid link overloading:
        \begin{equation}
        \label{eq:bw}
        \begin{array}{cc}
            \forall j,k \in N: & \sum_{i,h \in S}(x_{ij} \cdot x_{hk} \cdot rbw_{ih}) \leqslant (fbw_{jk} - bwTh) \\
        \end{array}
        \end{equation}
\end{itemize}

\noindent \cref{eq:lat} and \cref{eq:bw} contain the product of two binary variables $x_{ij}$, which implies that our considered problem is a Mixed Integer Quadratic Problem (MIQP). 
We can linearise it into a MILP by exploiting new binary variables $ y_{ijhk} = x_{ij} \cdot x_{hk}$ (i.e. the logical AND between $x_{ij}$ and $x_{hk}$) and the following additional constraints on them:
\begin{equation*}
\begin{array}{llll}
    y_{ijhk} & \leqslant & x_{ij} & \forall i \in S, j \in N\\
    y_{ijhk} & \leqslant & x_{hk} & \forall h \in S, j \in N \\
    y_{ijhk} & \geqslant & x_{ij} + x_{hk} - 1 & \forall i,h \in S, j,k \in N \\
    \\
    y \in \{0, 1\} & & &
\end{array}
\end{equation*}

\noindent The first two inequalities require the variable $y_{ijhk} = 0$ when at least one of the other two variables is 0. The third imposes $y_{ijhk} = 1 $ if $x_{ij} = x_{hk} = 1$.

\noindent The goal of our placement strategy is to find the best assignment of \textit{S} to \textit{N}, in order to minimise the overall placement cost. This means finding an assignment for the variables in \textit{X} which minimises \cref{eq:objective_function}, knowing the cost matrix \textit{C}. Summing up, the considered problem can be modelled as follows:

{\small
\begin{equation*}
\begin{array}{rcllcr}
    \min & \multicolumn{2}{l}{\sum_{i,j} c_{ij} \cdot x_{ij}} & \\
    \textrm{subj. to:} \\
    \sum_{j}x_{ij} & = & 1 & \forall i \in S & \\
    x_{ij} & \leqslant & b_j & \forall i \in S, j \in N \\ \\
    \sum_{j}b_{j} & \leqslant & MAX\_BIN & \forall j \in N & \\
    \sum_{i}(x_{ij} \cdot rhw_i) & \leqslant & (fhw_j - hwTh) & \forall j \in N \\
    \sum_{i,h}(y_{ijhk} \cdot rbw_{ih}) & \leqslant & (fbw_{jk} - bwTh) & \forall j,k \in N & \\
    y_{ijhk} \cdot flat_{jk} & \leqslant & rlat_{ih} & \forall i,h \in S, \forall j,k \in N \\
    \\
    x, y, b \in \{0, 1\} &
\end{array}
\label{eq:model}
\end{equation*}
}
\subsection{Declarative Preprocessing}
\label{sec:prolog_preprocessing}

The declarative preprocessing step, implemented in Prolog, involves two primary tasks, shown in \cref{fig:cr}:

\begin{enumerate}
    \item It performs \textit{continuous reasoning}~\cite{forti2022declarative,massa2022daplacer}, trying to reuse existing placements if they are still valid, thus minimising unnecessary recomputation and service migrations.
    \item It computes \textit{compatible nodes} for each application component (service or function), verifying the capabilities and requirements of the nodes.
\end{enumerate}

\begin{figure*}
\begin{codeNum}[firstnumber=1]
    preprocess(App, Compatibles) :-
        \plnot deployed(App, _),
        application(App, Functions, Services), 
        append(Functions, Services, Components),
        findCompatible(Components, Compatibles).
    preprocess(App, Compatibles) :-
        deployed(App, Placement),
        crStep(Placement, Compatibles).

    findCompatible([C|Cs], [(C,Compatibles)|Rest]):- compatiblePlacements(C,Compatibles), findCompatible(Cs, Rest).
    findCompatible([],[]).

    compatiblePlacements(C, SCompatibles):-
        findall((N, Cost), componentPlacement(C, N, [], Cost), [Comp|Atibles]), \textcolor{CadetBlue}{% [Comp|Atibles] is not empty}
        sort([Comp|Atibles], SCompatibles).
    
    crStep(P, Compatibles) :- crStep(P, [], Compatibles).
    crStep([(C,N)|P], POk, [(C,[(N,Cost)])|Cs]) :-
        componentPlacement(C, N, P, Cost), qosOK(C, N, POk),
        crStep(P, [(C,N)|POk], Cs).
    crStep([(C,N)|P], POk, [(C,Compatibles)|Cs]) :-
        \plnot ( componentPlacement(C, N, P, _), qosOK(C, N, POk) ), 
        compatiblePlacements(C, Compatibles),
        crStep(P, POk, Cs).
    crStep([], _, []).
    
    componentPlacement(F, N, P, Cost) :-
        functionInstance(F, FId, _), function(FId, SWPlat, (Arch,HWReqs)),
        node(N, SWCaps, (Arch,HWCaps), _, _), 
        requirements(FId, N), 
        member(SWPlat, SWCaps),
        hwOK(N, HWCaps, HWReqs, P),
        nodeType(N, Type), cost(Type, C, Cost).
    componentPlacement(S, N, P, Cost) :-
        serviceInstance(S, SId), service(SId, SWReqs, (Arch,HWReqs)),
        node(N, SWCaps, (Arch,HWCaps), _, _), 
        requirements(SId, N), 
        subset(SWReqs, SWCaps),
        hwOK(N, HWCaps, HWReqs, P),
        nodeType(N, Type), cost(Type, C, Cost).

    hwOK(N, HWCaps, HWReqs, P) :-
        findall(HW, hwOnN(N, P, HW), HWs), sum_list(HWs,TotHW),
        hwTh(T), HWCaps >= TotHW + HWReqs + T.

    qosOK(C, N, P) :- 
        findall(DF, relevant(C, N, P, DF), DataFlows), 
        checkDF(DataFlows, [(C,N)|P]).
    
    checkDF([(N1,N2, ReqLat, _, SecReqs)|DFs], [(C,N)|P]) :-
        secOK(N1, N2, SecReqs),
        link(N1, N2, FeatLat, FeatBW),
        FeatLat =< ReqLat,
        bwOK(C, N, FeatBW, P),
        checkDF(DFs, [(C,N)|P]).
    checkDF([], _).
    
\end{codeNum}
\caption{Declarative preprocessing step with continuous reasoning.}
\label{fig:cr}
\end{figure*}

The preprocessing phase starts with the predicate \cd{preprocess/2} (lines 1--8), which handles the selection of compatible placement candidates based on whether the application \cd{App} is currently deployed. If the application is not yet placed on the infrastructure (lines 1--5), the list of functions and services is collected and merged into a single component list and passed to \cd{findCompatible/2} (line 5) to compute compatible nodes. If a previous placement exists (line 7), the system switches to the continuous reasoning procedure via \cd{crStep/2} (line 8).

The compatibility assessment is performed by \cd{findCompatible/2} (lines 9--10), which iteratively invokes \cd{compatiblePlacements/2} for each component (lines 11--13). This predicate uses \cd{findall/3} to collect all nodes \cd{N} that satisfy the constraints for component \cd{C}, according to the predicate \cd{componentPlacement/4}.

The core predicate \cd{componentPlacement/4} (lines 23--36) checks whether a component \cd{C} (either a function or a service instance) can be hosted on a node \cd{N} under partial placement \cd{P}. The checks include:
\begin{itemize}
    \item \textit{architecture} and \textit{software} compatibility (\cd{member/2} or \cd{subset/2}),
    \item enforcement of non-numeric and/or component-specific deployment requirements via \cd{requirements/2},
    \item \textit{hardware} capacity verification through \cd{hwOK/4},
    \item and \textit{cost} computation using \cd{cost/3}.
\end{itemize}

Hardware availability is verified through \cd{hwOK/4} (lines 37--39), which ensures that the sum of current allocations on node \cd{N} plus the demand of the new component does not exceed the capacity of the node, leaving a safe margin defined by \cd{hwTh/1}.

The QoS constraints are enforced by the \cd{qosOK/3} predicate (lines 40--42). Collects all data flows involving the component \cd{C} and the candidate node \cd{N}, and delegates their validation to \cd{checkDF/2}.

The predicate \cd{checkDF/2} (lines 43--49) recursively evaluates each flow. It ensures that
\begin{itemize}
    \item \textit{security} constraints are respected via \cd{secOK/3},
    \item link featured \textit{latency} is lower than the required one,
    \item link available \textit{bandwidth} is sufficient, verified by \cd{bwOK/4}.
\end{itemize}

In case the application is already deployed (line 6), \cd{preprocess/2} delegates the control to the \cd{crStep/2} predicate (lines 14--22), which implements the core logic of continuous reasoning. This mechanism attempts to retain previously computed placements by verifying whether each component \cd{C} can still run on its assigned node \cd{N}, given the current state and deployment of the infrastructure (line 15). The validity of each retained placement is assessed using the \cd{componentPlacement/4} and \cd{qosOK/3} predicates, which collectively enforce both deployment and QoS constraints. If the assignment is still valid, the placement is preserved and its cost is recomputed (line 16); otherwise, the system recomputes the compatible nodes for that component via \cd{compatiblePlacements/2} (line 20). 

However, the continuous reasoning step is not guaranteed to succeed in restoring a complete placement: if at least one component cannot be reassigned to any node, the process is aborted, and the standard preprocessing routine is reapplied from scratch using \cd{findCompatible/2}.

\subsection{Pipeline Overview}
\label{sec:pipeline-overview}

\begin{figure}[htb]
    \centering
    \includegraphics[width=\linewidth, trim={4cm 5cm 4cm 5cm}, clip]{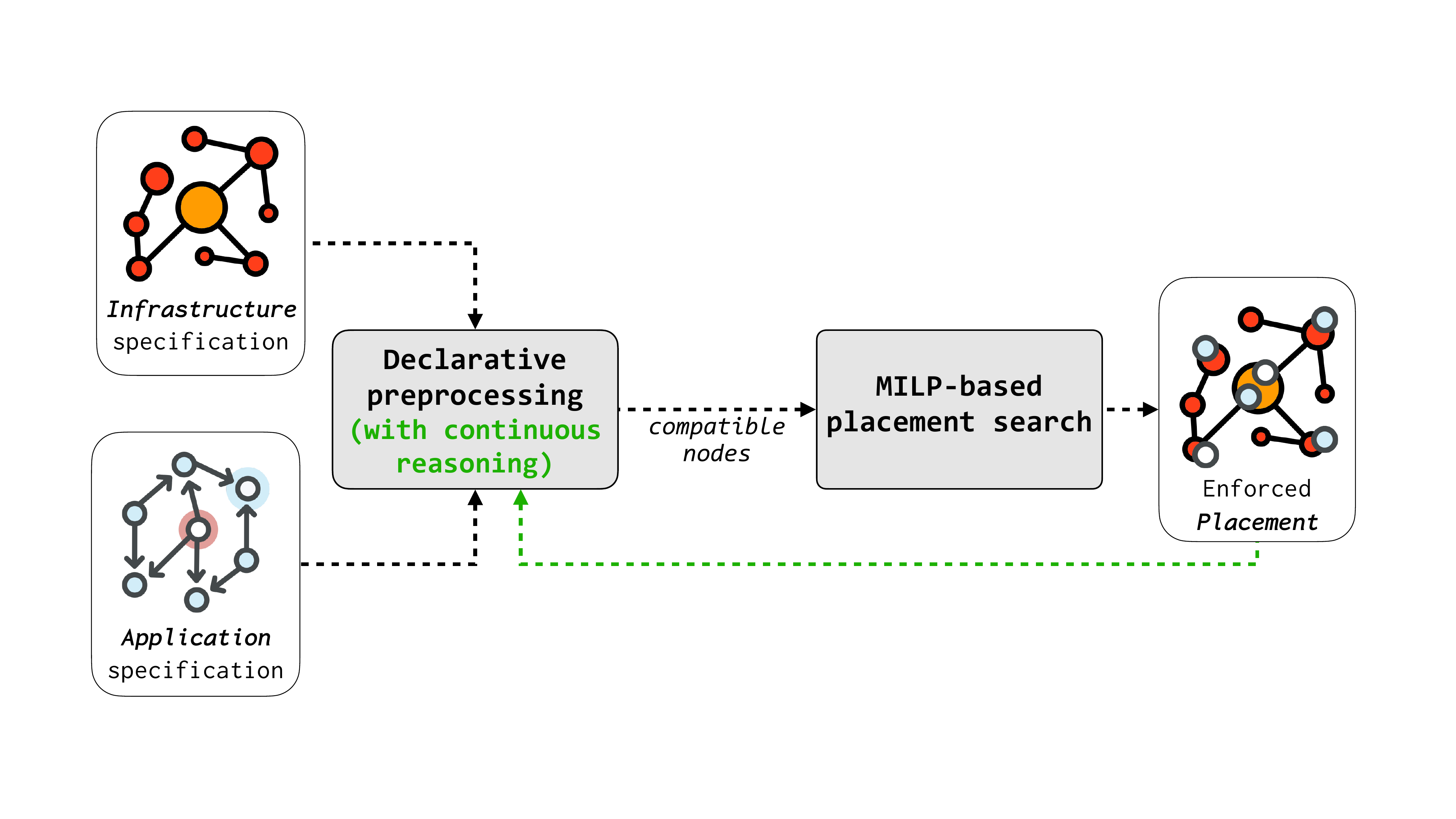}
    \caption{Overview of the \edgewise pipeline, consisting of a declarative preprocessing stage and a MILP-based placement search phase. With \edgewisecr we introduced \textit{continuous reasoning} in the first stage, and the reuse of previously found eligible placements.}
    \label{fig:workflow}
\end{figure}

\Cref{fig:workflow} illustrates the two-stage pipeline underlying \edgewisecr. The approach takes as input a multi-component application, specified in terms of qualitative and quantitative requirements, and the current infrastructure state. Then it applies two stages:

\begin{enumerate}
    \item \textit{Declarative preprocessing}: the first stage, implemented in Prolog, expresses and enforces non-numerical requirements such as software dependencies, architecture compatibility, security constraints, and geographical placement policies. For each component of the application (service or function), it produces a filtered set of \textit{compatible} infrastructure nodes, each annotated with an estimated provisioning \cd{cost}. This set is later used to reduce the complexity of the search space. In the extended version of \edgewisecr, this step also implements \textit{continuous reasoning}, enabling the reuse of existing placements if still valid. If the placement is no longer valid, it recomputes compatibility sets as in the baseline \edgewise.
    
    \item \textit{MILP-based placement search}: the second phase uses the output of the declarative stage to define and solve a resized MILP placement problem. Only variables related to compatible nodes are introduced into the optimisation model, significantly reducing the number of decision variables and constraints. This leads to shorter execution times for the model generation and solution phases. The output of this step is an enforced placement, which assigns each application component to one of the compatible nodes in a way that minimises overall provisioning cost.
\end{enumerate}

\section{Experimental Evaluation}
\label{sec:experiments}

\subsection{Experimental Setup}
\label{sec:experimental_setup}

To evaluate the proposed approach, we use the ECLYPSE simulator~\cite{massa2025eclypse}, which allows us to model Cloud-Edge infrastructures and simulate dynamic failures over time. The simulation spans 30 ticks, with each tick having a timeout of 1200 seconds (20 minutes). An \textit{update policy} is applied at each tick, introducing failures by making 10\% of the infrastructure nodes unavailable.

The involved infrastructures vary in size, containing $2^i$ nodes, with $i \in [6,11]$. Each node is assigned a random type among \cd{cloud}, \cd{edge}, and \cd{thing} and is characterised by its hardware capabilities and software configurations. The available hardware resources are determined by sampling a normal distribution and then truncated to the nearest integer, with values trimmed to the range $[32, 1024]$ to ensure consistency. The network is configured with bandwidth values randomly selected in $[20,500]$ Mbps, while initial latencies are chosen within $[2, 20]$ ms. Then, we redefine latencies using the Floyd-Warshall algorithm~\cite{hougardy2010floyd} to compute \textit{all-pairs shortest paths}.

To study the impact of different connectivity patterns, we generate infrastructures using three distinct network topology models: 
\begin{enumerate*}[label=\textit{(\roman{*})}]
    \item the Barabasi-Albert model (BA), which produces scale-free networks with highly connected hubs; 
    \item the Erdos-Renyi model (ER), representing random graphs with uniform connectivity probability, and 
    \item an Internet-as-a-graph topology (IAG), inspired by autonomous system graphs.
\end{enumerate*}

The experiments consider three multi-service applications with different computational requirements, dependencies, and data flow constraints\footnote{The complete knowledge base representing the applications is available at: \url{https://github.com/di-unipi-socc/edgewise/tree/cr/edgewise/applications/prolog}}. \cd{speakToMe} is the application presented in \cref{sec:problem}; \cd{arFarming} is a precision farming application integrating augmented reality and IoT to facilitate real-time crop monitoring, adapted from~\textcite{phupattanasilp2019ariot}. Finally, \cd{distSecurity}, based on~\textcite{petiwala2021smart}, is a distributed security application designed for smart parking systems, incorporating automated license plate recognition to enhance security and parking automation.

To assess the performance of the different reasoning strategies, each experiment is executed under three distinct planning configurations: 
\begin{enumerate*}[label=\textit{(\roman{*})}]
    \item a purely declarative approach, relying on our prior Prolog solution~\cite{edgewise} to compute a valid deployment that satisfies all constraints without performing any explicit optimisation;
    \item the \edgewise planner~\cite{edgewise}, which extends the Prolog-based pre-process filtering step with a MILP formulation to compute cost-optimal placements, following the model described in~\cref{sec:milp_formulation}, and 
    \item \edgewisecr, which incorporates continuous reasoning by introducing a lightweight heuristic to reduce problem size over time. 
\end{enumerate*}

The experimental evaluation focuses on three key metrics: the \textit{placement cost}, computed according to the declarative model in \cref{sec:prolog_formulation} while considering infrastructure and application constraints, the \textit{execution time} required to determine a placement solution, and the number of migrated services at each tick of the simulation. Each experiment is executed three times using fixed random seeds to ensure reproducibility\footnote{The complete experiment code is available at: \url{https://github.com/di-unipi-socc/edgewise/tree/cr}}.
\subsection{Discussion of results}
\label{sec:results}

\begin{figure*}[!ht]
  \centering

  \begin{subfigure}[t]{0.32\textwidth}
    \centering
    \includegraphics[width=\linewidth]{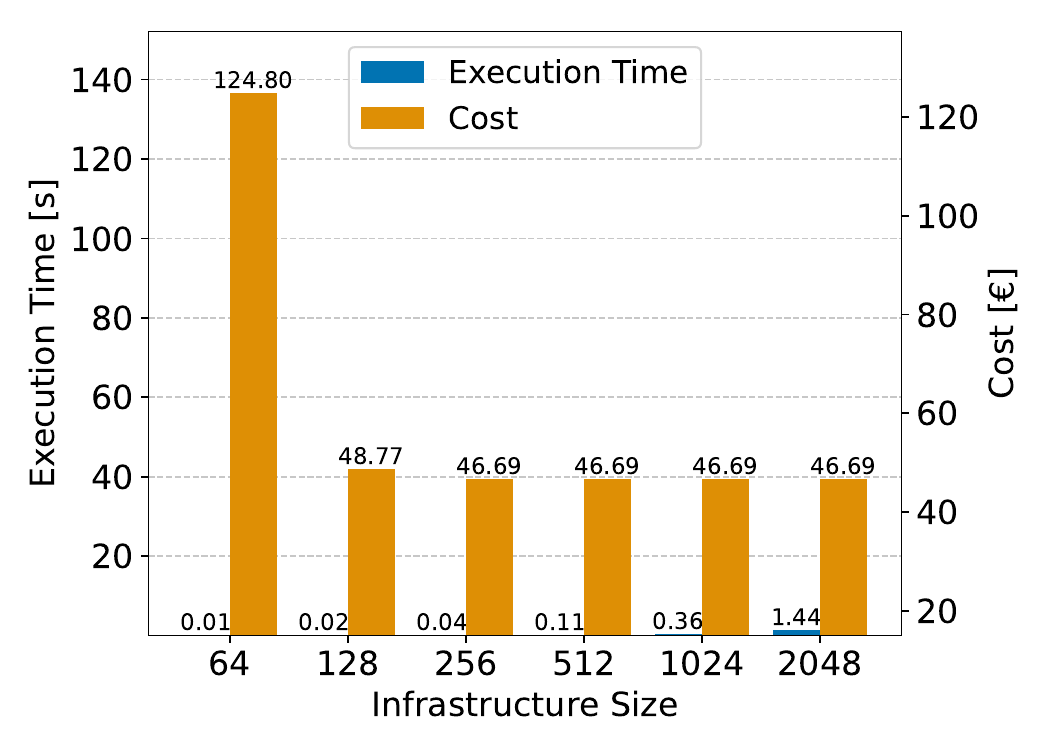}
    \caption{\fncd{arFarming} -- \textit{prolog}}
    \label{fig:arfarming-prolog}
  \end{subfigure}
  \begin{subfigure}[t]{0.32\textwidth}
    \centering
    \includegraphics[width=\linewidth]{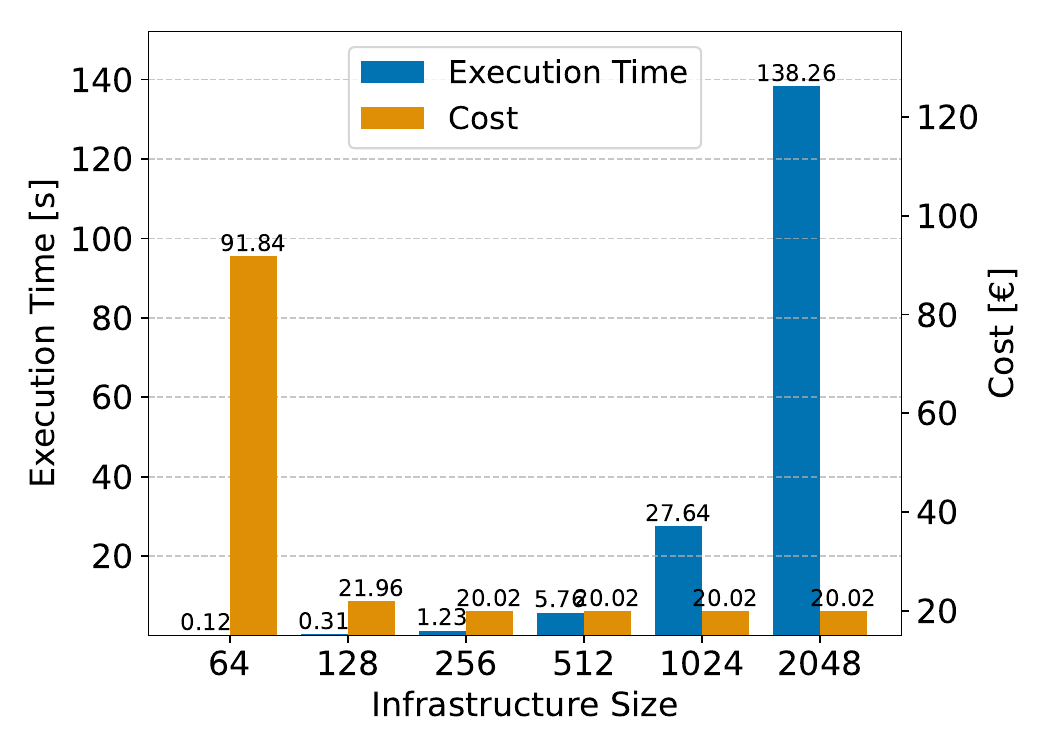}
    \caption{\fncd{arFarming} -- \textit{\edgewise}}
    \label{fig:arfarming-edgewise}
  \end{subfigure}
  \begin{subfigure}[t]{0.32\textwidth}
    \centering
    \includegraphics[width=\linewidth]{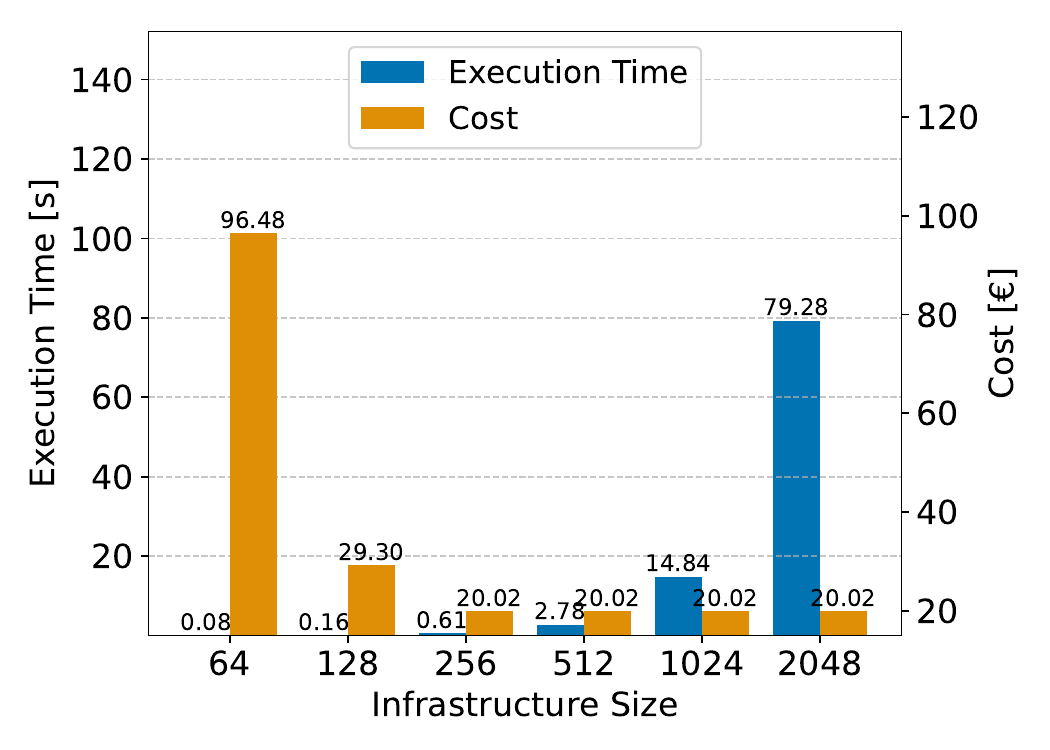}
    \caption{\fncd{arFarming} -- \textit{\edgewisecr}}
    \label{fig:arfarming-cr}
  \end{subfigure}

  \begin{subfigure}[t]{0.32\textwidth}
    \centering
    \includegraphics[width=\linewidth]{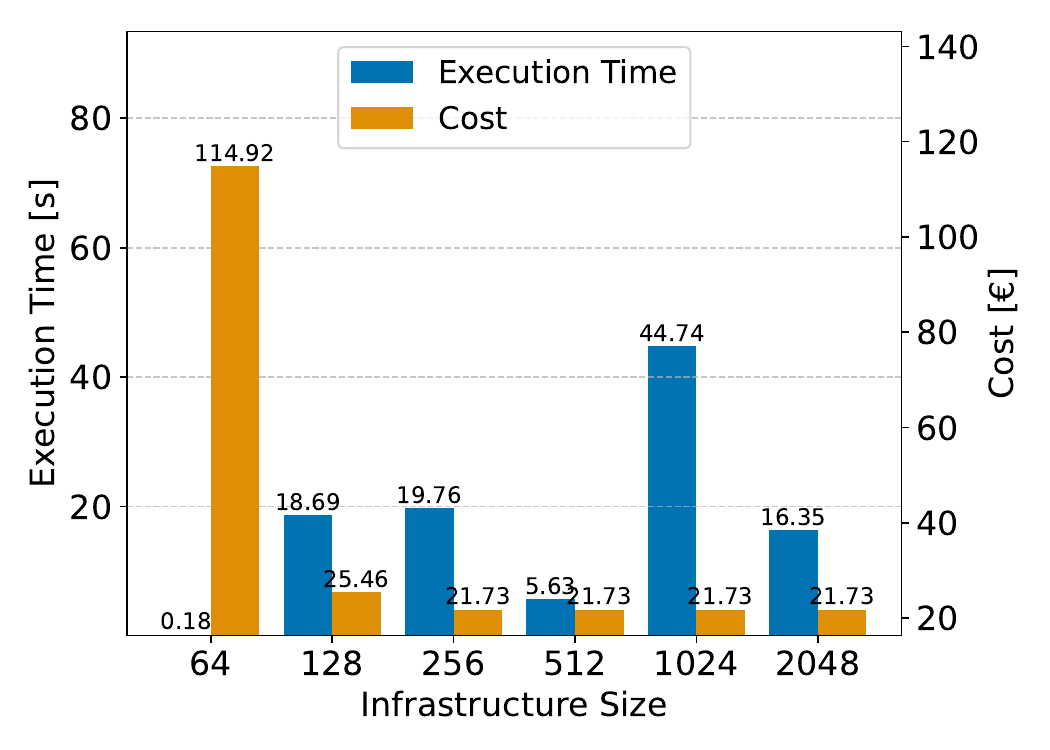}
    \caption{\fncd{distSecurity} -- \textit{prolog}}
    \label{fig:distsecurity-prolog}
  \end{subfigure}
  \begin{subfigure}[t]{0.32\textwidth}
    \centering
    \includegraphics[width=\linewidth]{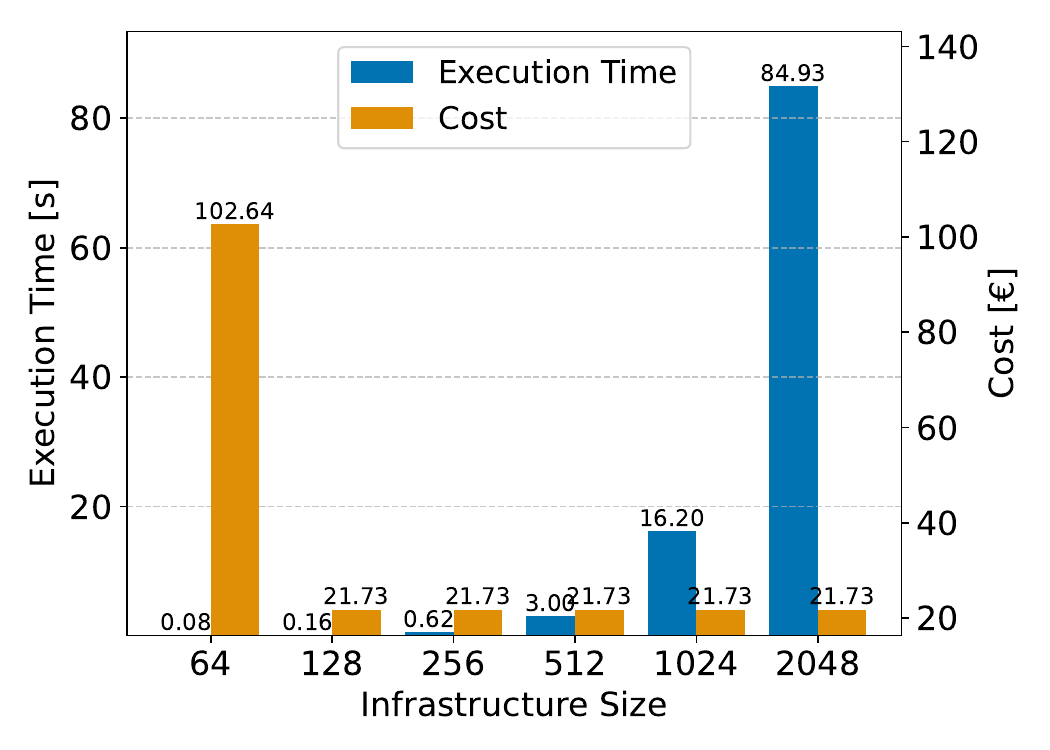}
    \caption{\fncd{distSecurity} -- \textit{\edgewise}}
    \label{fig:distsecurity-edgewise}
  \end{subfigure}
  \begin{subfigure}[t]{0.32\textwidth}
    \centering
    \includegraphics[width=\linewidth]{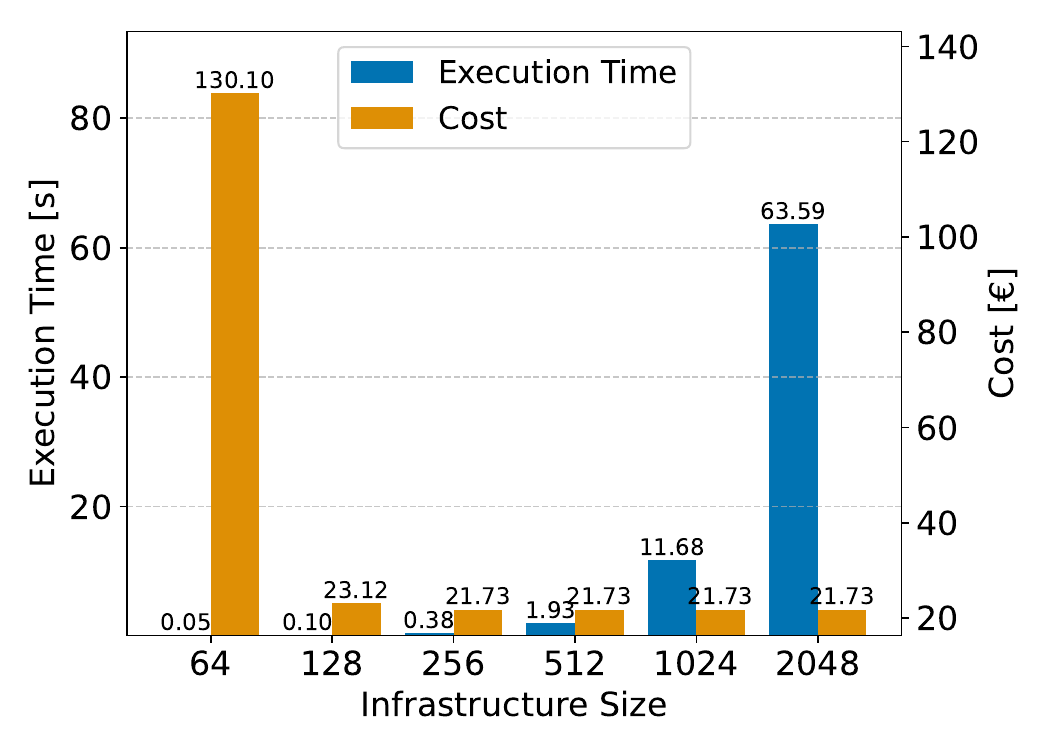}
    \caption{\fncd{distSecurity} -- \textit{\edgewisecr}}
    \label{fig:distsecurity-cr}
  \end{subfigure}

  \begin{subfigure}[t]{0.32\textwidth}
    \centering
    \includegraphics[width=\linewidth]{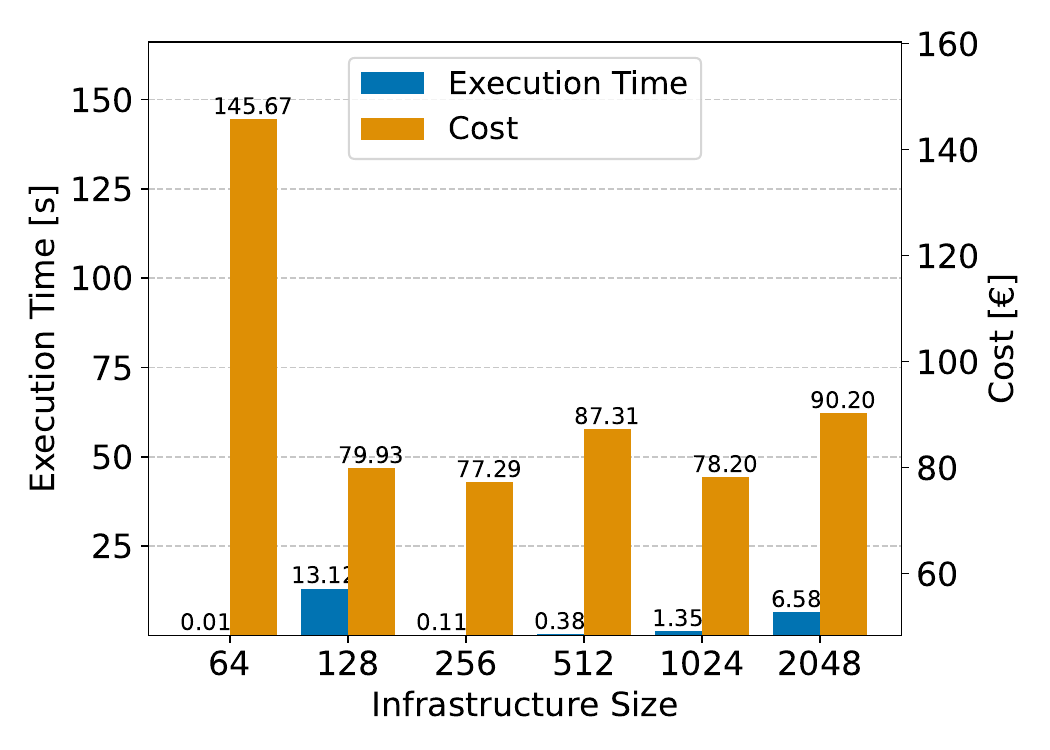}
    \caption{\fncd{speakToMe} -- \textit{prolog}}
    \label{fig:speakToMe-prolog}
  \end{subfigure}
  \begin{subfigure}[t]{0.32\textwidth}
    \centering
    \includegraphics[width=\linewidth]{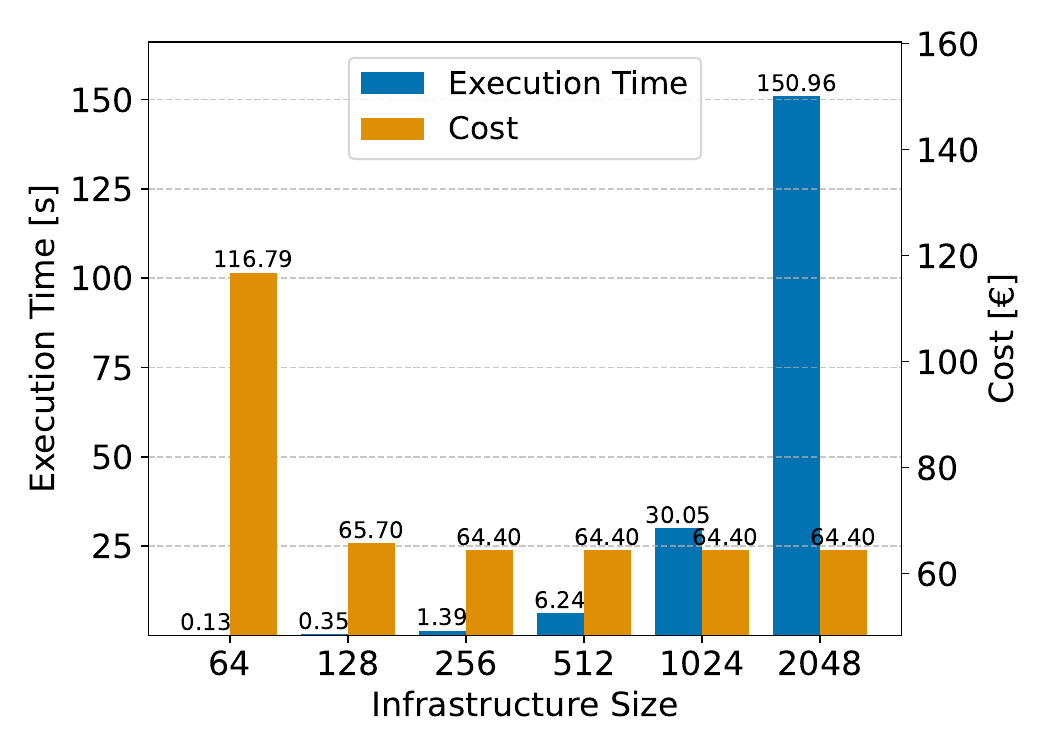}
    \caption{\fncd{speakToMe} -- \textit{\edgewise}}
    \label{fig:speakToMe-edgewise}
  \end{subfigure}
  \begin{subfigure}[t]{0.32\textwidth}
    \centering
    \includegraphics[width=\linewidth]{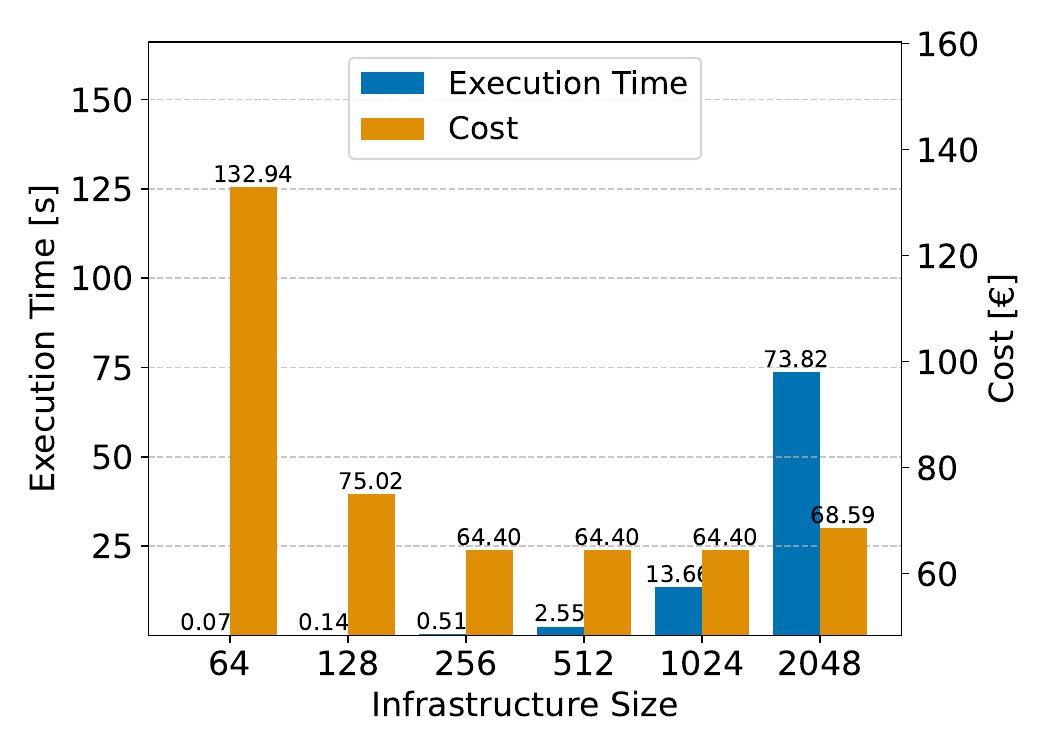}
    \caption{\fncd{speakToMe} -- \textit{\edgewisecr}}
    \label{fig:speakToMe-cr}
  \end{subfigure}

  \caption{\textit{Execution times} and \textit{deployment costs} across the three scenarios for each reasoning strategy. Each row corresponds to one application, and each column to a planning version.}
  \label{fig:time-cost-all}
\end{figure*}

The experimental results across the three application scenarios (\textit{arFarming}, \textit{distSecurity}, and \textit{speakToMe}) highlight consistent trends related to scalability and performance while also showcasing the benefits introduced by the continuous reasoning mechanism.


As expected, execution time increases with the size of the topology. In the case of \edgewise, this growth follows an exponential trend, as clearly visible in \cref{fig:time-cost-all}. This behaviour can be attributed to the MILP solver, which, although supported by a preliminary filtering phase performed in Prolog, still needs to process a constraint matrix that becomes increasingly large and populated as the infrastructure grows. Specifically, the number of decision variables and constraints increases with the number of available nodes and deployable components, making the MILP problem more computationally intensive and harder to solve, causing a sharp increase in execution time.

In contrast, the Prolog-based planner consistently shows very low execution times, even as the infrastructure scales. Nevertheless, some irregularities emerge in specific domains, particularly in \textit{distSecurity} (\cref{fig:distsecurity-prolog}), where execution times are more erratic and occasionally increase in a non-linear or exponential fashion. This suggests that certain application domains exhibit more complex search spaces, which influence the typically steady performance of Prolog.

The introduction of continuous reasoning in \edgewisecr leads to substantial improvements in execution time. Across all scenarios, we observe reductions ranging from 25\% to 65\% compared to standard \edgewise (\cref{fig:arfarming-cr,fig:distsecurity-cr,fig:speakToMe-cr}). These improvements are more significant in larger topologies, where the MILP formulation would otherwise dominate runtime.

About provisioning cost, an increase in infrastructure size tends to reduce variability. With more available nodes -- often belonging to lower-cost layers -- solutions tend to have similar or identical provisioning costs. This effect reflects the abundance of equivalent low-cost alternatives in large infrastructures, even if the cost model is relatively simple.

\begin{table}[ht]
\centering
\begin{tabular}{|c|c|c|c|}
\hline
\textbf{\# Nodes} & \textbf{arFarming} & \textbf{distSecurity} & \textbf{speakToMe} \\
\hline
64   & 35\% & 37\% & 38\% \\ \hline
128  & 44\% & 38\% & 36\% \\ \hline
256  & 43\% & 39\% & 38\% \\ \hline
512  & 39\% & 41\% & 37\% \\ \hline
1024 & 34\% & 40\% & 38\% \\ \hline
2048 & 35\% & 39\% & 38\% \\ \hline
\end{tabular}
\caption{Average reduction in the number of migrated services with continuous reasoning (\edgewisecr) compared to plain recomputation (\edgewise), across different infrastructure sizes.}
\label{tab:migrated-services}
\end{table}

Finally, we analyse the number of migrated services per tick, which further illustrates the stabilising effect of continuous reasoning. As shown in \cref{tab:migrated-services}, \edgewisecr reduces the number of service relocations by approximately 35\% to 44\% in \cd{arFarming}, 37\% to 41\% in \cd{distSecurity}, and around 36\% to 38\% in \cd{speakToMe}, depending on the size of the infrastructure. These consistent reductions indicate that \edgewisecr effectively reuses existing valid placements, limiting unnecessary reconfigurations.

Notably, \edgewisecr can maintain the current placement entirely, without requiring any service migration, in 55\% of the simulation ticks. In contrast, \edgewise -- which recomputes a new solution from scratch at each tick -- results in zero migrations only 33\% of the time. This highlights the advantage of our continuous reasoning mechanism: it avoids redundant reallocation when possible and increases placement stability over time.
The benefit remains relatively stable even as the infrastructure grows. This suggests that while a larger infrastructure reduces the likelihood of critical node failures, the ability of \edgewisecr to preserve placements remains advantageous across scales.

Each method reveals a specific trade-off between execution time and provisioning cost:

\begin{itemize}

    \item The Prolog-based approach excels in response time, returning feasible solutions in milliseconds. However, this comes at the expense of deployment quality, since the cost associated with the declarative strategy can be up to 50\% higher than the optimal found by the MILP one.
    \item In contrast, \edgewise produces near-optimal deployments, but suffers from longer execution times, particularly on large infrastructures. This trade-off makes it suitable for contexts where solution quality is critical and longer planning times are acceptable.
    \item \edgewisecr strikes a valuable balance between these two approaches. Although its cost may increase -- up to 33\% in some scenarios -- it still provides significantly faster execution than standard \edgewise. The increase in cost is primarily due to the continuous reasoning heuristic constraints: some services are fixed onto specific nodes to simplify the MILP model and reduce the solving time and the number of migrations.
    
\end{itemize}

A comparative analysis of the three approaches -- Prolog-only, \edgewise, and \edgewisecr -- reveals a nuanced trade-off among execution time, placement stability, and provisioning cost. Although fast because of its heuristic nature, the Prolog baseline offers suboptimal placements and lacks robustness in dynamic settings. \edgewise improves significantly by using an MILP solver, producing cost-optimal solutions at the expense of longer computational times, which limits its applicability in scenarios with frequent reconfiguration needs. \edgewisecr, instead, strikes a balance between the two extremes: introducing continuous reasoning drastically reduces the number of full re-computations while retaining optimality. When the rate of infrastructure changes or placement requests increases, \edgewisecr demonstrates greater adaptability and responsiveness. Its ability to preserve valid placements results in lower cumulative execution costs over time, making it particularly suited for real-world deployments where reactivity and efficiency must coexist.

To further assess the practical benefits of continuous reasoning, we consider a simplified performance model based on an M/M/1 queuing system~\cite{Vilaplana2014}, where placement requests arrive dynamically and are processed by a single planning engine. In this setting, incoming requests are dropped if the queue becomes saturated, reducing the infrastructure operator's profit. In the worst-case scenario -- placing the \cd{speakToMe} application on an infrastructure of 2048 nodes -- \edgewisecr can process nearly twice as many requests as \edgewise before saturation occurs, according to the monitored execution times. This trend, which holds for all tested applications and infrastructure sizes, implies that the provider can potentially double their revenue by adopting \edgewisecr, assuming a fixed profit per accepted request. This gain compensates for the increase in provisioning cost introduced by continuous reasoning, which can reach 33\% compared to the optimal placements computed by \edgewise.

\section{Related Work}
\label{sec:related_work}

The problem of optimising service placement in Cloud-Edge environments has gained significant attention in recent years. Various strategies have been proposed to enhance QoS and QoE in Edge computing systems~\cite{salaht2020overview,santoso2017dynamic,malazi2022dynamic}. Various methodologies have been explored for the placement and management of multiservice applications in Cloud-IoT infrastructures~\cite{brogi2019place,applicationmanagementfogsurvey}.

Many existing approaches rely on centralised cloud-based orchestration mechanisms for service placement. These solutions often deploy services closer to users to optimise performance while reducing latency~\cite{Carrusca19,ullah2021micado,TangAccess21,maia2019optimized,guoAccess19}. Some of these methods adopt decentralised and self-organising mechanisms to manage service distribution dynamically. In contrast, our work focuses on a hybrid declarative and MILP approach, enabling adaptive placement while incorporating continuous reasoning to limit unnecessary recomputation.

Several studies have explored data-aware service placement, aiming to optimise data movement and access latency. Strategies such as moving data closer to computation units have been widely discussed~\cite{10.1145/3366372,8548069, ferrucci2016multidimensional, payberah2013lightweight}. Data replication and placement models have also been proposed to improve system responsiveness~\cite{8244318,LI20191}. Recent works address the interaction between service placement and data availability, as presented by \textcite{farhadi2021service}, who jointly optimise service placement and request scheduling for data-intensive applications, and \textcite{xie2023data}, who propose an efficient data placement and retrieval framework for cooperative edge clouds.

Mobility-aware approaches have also been considered to optimise placement decisions. \textcite{kimovski2022mobility} propose a model that incorporates Edge device mobility predictions to ensure stable application placement. Similarly, \textcite{malazi2022dynamic} review dynamic service placement approaches in multi-access Edge computing (MEC), highlighting how mobility and dynamic reallocation impact system performance. Our approach builds upon these ideas, but extends them by introducing continuous reasoning to proactively reduce service migrations.

Security-aware placement mechanisms have been explored to balance computation offloading with security risks. \textcite{sun2023security} propose a security-aware, time-efficient task scheduling model that optimises risk-aware service placement. This aligns with our work, as our continuous reasoning approach ensures minimal unnecessary reallocation while maintaining placement security constraints.

Several works have explored MILP-based solutions for service placement in Edge-Cloud environments, focusing on different optimisation objectives. \textcite{han2022energyefficient} propose a MILP model for energy-efficient service placement by minimising resource fragmentation in mobile Edge clouds. Similarly, \textcite{asgarian2024servicefunction} developed a MILP-based approach for service function chaining in Industrial IoT, incorporating heuristic approximations for large-scale networks. A cost-aware MILP formulation for Edge server deployment is introduced by \textcite{shao2022costaware}, optimising the user experience while balancing deployment costs. Additionally, \textcite{islam2024hyperheuristic} present a hyper-heuristic MILP model that considers both service placement and QoE requirements in 5G Mobile Edge Computing. A broader survey by \textcite{malazi2022dynamic} reviews dynamic MILP-based service placement strategies, discussing their adaptability to real-time changes in Edge-Cloud infrastructures. Although these approaches demonstrate the flexibility of MILP for placement optimisation, they do not incorporate declarative reasoning and continuous adaptation mechanisms, which are central to our hybrid approach.

Declarative programming has been used for the placement of services considering the constraints of QoS, cost, and adaptation~\cite{forti2022declarative,forti2021declarative,forti2022green}. In addition, aggregate computing approaches have been proposed for distributed service coordination~\cite{casadei19,casadei21}. However, to the best of our knowledge, no prior work has combined declarative programming with MILP while incorporating continuous reasoning to minimise service migrations and placement overhead.


Continuous reasoning has been explored to improve runtime decision-making in Cloud-Edge placements. \textcite{DBLP:journals/fgcs/FortiGB21} introduced a Prolog-based strategy to select QoS- and context-aware placements, thus suffering when an optimal solution placement needs to be found. More recently, \textcite{DBLP:journals/computing/HerreraBFBM23,DBLP:journals/tsc/HerreraBFBM24} proposed a MILP-based layered heuristic to continuously adapt deployments. Lastly, \textcite{azzolini2024continuous} applied logic programming to distribute base container images across Cloud-Edge networks, addressing a complementary yet related challenge.

Overall, this work advances the state-of-the-art by integrating declarative programming and MILP-based optimisation within a continuous reasoning framework, ensuring optimal service placement while dynamically adapting to infrastructure changes.
\section{Conclusions}
\label{sec:conclusions}

This work extends our previous \edgewise methodology for multi-component application placement in Cloud-Edge environments by introducing a continuous reasoning mechanism. This enhancement allows reuse of previously computed placements, significantly reducing unnecessary re-calculations and service migrations, especially under dynamic infrastructure conditions. Combined with the declarative filtering stage and the MILP-based optimiser, this extension forms a flexible and adaptive pipeline capable of providing optimal placements with substantially lower execution times.

Our experimental evaluation, conducted in three realistic application scenarios and diverse infrastructure topologies, demonstrates the effectiveness of \edgewisecr. Compared to its predecessor, it achieves up to 65\% faster execution, with only a modest increase in placement cost, and consistently maintains placement stability. These results confirm continuous reasoning is a practical and efficient solution for real-time or frequently changing Cloud-Edge environments.

Future work will explore several directions. We plan to incorporate more expressive cost models that account for energy consumption and carbon footprint, thus extending our current provisioning focus towards sustainability-aware placement. Additionally, we envision enriching the decision-making logic with support for predictive models that anticipate failures or load variations, allowing proactive rather than reactive placement adaptation. Finally, we aim to test such solutions on real testbeds, validating their performance and adaptability in practical deployment scenarios.

\AtNextBibliography{\footnotesize}
\printbibliography

\end{document}